\documentclass[12pt]{article}
\usepackage{amsmath,amssymb,amsfonts}

\usepackage{geometry}                
\usepackage{subeqnarray,eqsection,indent,cite}

\usepackage{slashed}

\newcommand{\uh}{{\hat u}}
\newcommand{\vh}{{\hat v}}

\newcommand{\bone}{1\!\!1}

\newcommand{\tr}{\mathop{\rm tr}\nolimits}
\newcommand{\Tr}{\mathop{\rm Tr}\nolimits}

\newcommand{\pf}{\mathop{\rm pf}\nolimits}
\newcommand{\sgn}{\mathop{\rm sgn}\nolimits}

\newcommand{\fc}{{\cal F}}

\newcommand{\giro}[1]{\stackrel{\circ}{#1}}
\newcommand{\nc}{\giro{R}}
\newcommand{\mc}{R}
\newcommand{\nt}{\nc\hspace{-1mm}{}_t}
\newcommand{\mt}{\mc_t}
\newcommand{\me}{E}

\newcommand{\Iaa}{I}           

\newcommand{\BNd}{\giro{\cal F}_{N,\,t}}    
\newcommand{\NdB}{{\cal F}_{N,\,t}  }   
\newcommand{\BNdt}{\giro{\cal F}_{N,\,t+1}}   
\newcommand{\BNdtm}{\giro{\cal F}_{N,\,t-1}  } 

\newcommand{\be}{\begin{equation}}
\newcommand{\ee}{\end{equation}}

\newcommand{\reff}[1]{(\ref{#1})}

\begin{document}

\title{Diquarks in the nilpotency expansion
of QCD and their role at finite chemical potential}

\author{
  {\small Sergio Caracciolo}                       \\[-1.7mm]
  {\small\it Dipartimento di Fisica and INFN}      \\[-1.7mm]
  {\small\it Universit\`a degli Studi di Milano}   \\[-1.7mm]
  {\small\it via Celoria 16, I-20133 Milano, ITALY}                \\[-1.7mm]
  {\small\tt Sergio.Caracciolo@mi.infn.it},        \\[-1.7mm]
  {\protect\makebox[5in]{\quad}}
   \\
  {\small Fabrizio Palumbo}  \\[-1.7mm]
  {\small\it INFN -- Laboratori Nazionali di Frascati}      \\[-1.7mm]
  {\small\it P.~O.~Box 13, I-00044 Frascati, ITALY}       \\[-1.7mm]
   {\small\tt fabrizio.palumbo@lnf.infn.it}       \\[-1.7mm]
  \\
 {\protect\makebox[5in]{\quad}} }

\maketitle

\thispagestyle{empty}

\begin{abstract}

We assume that  the most important quark correlations are pairwise at all baryon densities.  We introduce correlated pairs by means of Bogoliubov transformations which are functions of   time and spatial gauge fields, in the formalism of the transfer matrix with lattice regularization. The dependence on time and gauge fields allows us to enforce  gauge invariance and other symmetries term by term in the transformed quantities. The resulting action should be suitable for the description of multiquark mesons and baryons as states of a quark and a diquark. We derive the quark  contribution to the free energy at finite chemical potential in a certain approximation. Its expression  cannot be evaluated analytically, but it  has a definite sign.

\end{abstract}

\clearpage

\vfill\eject

\renewcommand{\thefootnote}{\arabic{footnote}}
\setcounter{footnote}{0}

\clearpage

\section{Introduction}

Our understanding of QCD, at least in the nonperturbative regime for the gauge coupling constant, strongly relies on numerical simulations which, indeed, have become more and more performing over the years. It would be desirable, however, to construct from first principles an
approximate description suitable to capture efficiently what we believe are the essential features. And, possibly, 
 introduce a perturbative scheme (in some parameter) in order to improve systematically the initial approximation.

This kind of approach  becomes more appealing when we wish to investigate QCD at finite chemical potential, because the numerical simulations of the fermion sector are plagued by the infamous sign problem. For example, in the fermion contribution to the free energy, this difficulty is revealed by the large accuracy needed in the evaluation of terms which (almost) cancel out among themselves~\footnote{ For attempts to tackle this problem by analytic continuation in the chemical potential one can see~\cite{Phil,MP,V1,V2}.}.

We suspect that the fermion contributions affected by sign fluctuations are due to states of high energy. In such a case, since these contributions to the free energy (almost) cancel out,  an  approximation which retains  only the more stable fermion states, for any gauge field configuration, would reasonably solve the problem, simply because it neglects these fluctuations altogether.

There are several indications which might help in selecting such fermion states. 
At low baryon density, many authors think that {\em diquarks} are important substructures in hadrons. Actually, in a historical perspective, baryonic constituents with diquark quantum numbers were already hypothesized by Gell-Mann~\cite{Gell}, but as elementary constituents. Later, with various motivations, models of baryons constructed in terms of one quark and one diquark have been investigated~\cite{Ida,Lic1,Lic2,Zralek,Jaffe,Hess,Chaka, Sriv}, 
and the diquark was regarded as a really composite state, even though, in practice, it was often treated as elementary. 
Subsequently, diquarks played an increasingly relevant role in the interpretation and explanation of several properties of mesons~\cite{Rose,Hendry,Cahi,Maia,Polosa}. 

Also at very high baryon density  diquarks  appear of fundamental importance as the various phases of color superconductivity can be understood in terms of their condensation~\cite{Barr,Bailin,Alford,Rapp}. 

A relatively smaller amount of work has been done at intermediate densities, where, however, it  has been suggested that the structure of condensed diquarks might change with decreasing baryon density, while their size might shrink down to a dimension comparable with the average interquark distance.  And this could explain the crossover from color superconductivity to Bose-Einstein condensation of molecular diquarks~\cite{Mats,Abuki,Kitazawa,Mishra}.

In conclusion, we deem that the results reported above~\footnote{The quoted references are only a sample of a vast literature, chosen to support our arguments. We apologize to Authors whose important contributions have not been acknowledged here.}
 strongly suggest that two-quark correlations are very important at all baryon densities. In the following we shall take seriously this hint  and we will assume that fermion states which are energetically stable, should always contain diquarks, in a condensed phase at high baryon density,  accompanied by unpaired quarks at low baryon density. 

We must emphasize that when we talk about a diquark we mean only a pair of correlated quarks.  Therefore, the simple presence of diquarks does not necessarily imply the existence of real or virtual bound states of two quarks, as the molecular diquarks  of a Bose-Einstein condensate or the Cooper pairs of  a color  superconductive phase.  Only above some critical values of the chemical potential such states can, eventually, appear. 

At a formal level, we will adopt the lattice regularisation and use the Kogut-Susskind formulation for fermions. 

As a first step, we need to identify positive and negative energy states of the Dirac lattice Hamiltonian. For this purpose we perform a first Bogoliubov transformation, which is equivalent to a Foldy-Wouthuysen transformation~\cite{CPV}.
Bare fermions are replaced by quasiparticles in presence of a background field. This construction is really meaningful only for a class of gauge configurations that we shall call {\em stationary}  meaning that the corresponding chromomagnetic field is independent on time, and the chromoelectric field is vanishing~\cite{noi-prd}. We remark that these conditions are  well suited to  study also the effect of an intense background magnetic field on strong interactions, a problem considered of interest both at the level of the cosmological electroweak phase transition and for the heavy-ion collisions. And, indeed, numerical simulations have already been performed both in the quenched approximation (see, for example,~\cite{Misha}) and with dynamical fermions (see~\cite{Negro} which contains also a detailed bibliography), in order to try to understand magnetic catalysis, i.e. the increase of chiral symmetry breaking induced by the magnetic background field.

Afterwards, in this paper, we introduce diquarks  by means of a second Bogoliubov transformation, 
 so that diquarks will appear as {\em Cooper pairs of quasiparticles}.

In the fermion contribution to the free energy, at {\em fixed stationary} gauge-field configuration, we can distinguish, in our formalism, a contribution ${\cal S}_{bo}$ from the vacuum, which we call  {\em bosonic}, 
from the fermonic action of the remaining  quasiparticles. 
At low temperature ${\cal S}_{bo}$ is dominating and becomes {\em exactly} the whole free energy at zero temperature. At vanishing chemical potential $\mu$, it takes the form
\be
{\cal S}_{bo} =  - \frac{L_0}{2} \tr \ln \overline{Q} = - \frac{L_0}{2}   \sum_i \ln Q_i \label{primaA}
\ee
where $L_0$ is the size in the temporal direction of the lattice, and is therefore the inverse temperature, and $L_0/2$ is the number of blocks in which we have to divide the time-direction. The operator $\overline{Q}$ is related, as we shall see later, to the operator $N$ (see its explicit form in~\reff{Dirac}) appearing in the definition of the transfer matrix of the quarks, and the $Q_i$ (respectively $N_i$) are its gauge-independent eigenvalues corresponding to eigenstates that we enumerate by using the index $i$. More precisely
\be
Q_i = 1 + \frac{1}{2} \left[ N_i^\dagger N_i + \sqrt{\left( N_i^\dagger N_i  \right)^2 + 4 N_i^\dagger N_i } \right] \label{qi}
\ee
For all states $i$ these eigenvalues are, by inspection, real and larger than unity, that is $\ln Q_i \ge 0$.  The result~\reff{primaA} coincides with what we got without the introduction of diquarks~\cite{noi-prd}.

Let us now introduce a positive chemical potential. We find that ${\cal S}_{bo}$ decreases according to:
\be
{\cal S}_{bo} = \frac{L_0}{2} \left\{ - \sum_i \ln Q_i - \sum_{i\in P} \left[ 2 \mu - \ln Q_i\right] \right\}\, , \label{sbo}
\ee
where the set $P$ denotes quasiparticles states $i$ such that $\ln Q_i <  2\mu $.
The above estimate has been derived by assuming that diquarks are formed by quasiparticles in $P$ with a  {\em simple pairing} structure, namely,  for each $i\in P$ there is one and only one conjugate state $p(i)\in P$, with $Q_{p(i)}=Q_i$. 

As the eigenvalues $Q_i$'s do not depend on the chemical potential,  the fermonic number $n_F$, defined by
\be
n_F = - \frac{1}{L_0} \frac{\partial {\cal S}_{bo} }{\partial \mu}\, ,
\ee
exactly counts the number of paired quark states of the ensemble $P$. Thanks to the introduction of diquarks we have taken into account in the bosonic action the nonvanishing fermionic number.
At increasing chemical potential the background field is depleted because of Pauli blocking. 
A short account of this analysis has already been presented in~\cite{Lattice2010}.

To derive~\reff{sbo} we computed the bosonic contribution to the vacuum after the two Bogoliubov transformations which introduce, respectively, the background and the diquark field. If, on one hand, \reff{sbo} is recovered by the fields which satisfy a variational principle, the general expression, on the other hand, can be used to study multi-quark mesons and baryons as quark-diquark composites by taking into account the  fluctuations of the background and diquark fields, along the lines of the expansion presented in~\cite{CLP}.

For each fixed stationary gauge-field configuration we get from the condition $\log Q_i < 2 \mu$ a sharp Fermi surface. After the integration on this space of  gauge-field configurations we expect that the Fermi surface should be smoothed out. This is confirmed by a perturbative calculation in the gauge coupling-constant that we performed
for large values of the chemical potential. In this case we get also a gap equation compatible with that obtained by standard methods~\cite{Kogut}.

We think it is useful to compare our results with the nonrelativistic ones.
From the technical point of view our formalism is a fermion number conserving extension of the  theory of superconductivity developed by Bogoliubov and Valatin~\cite{Bogo,Valatin} which violates this symmetry. The enforcement of fermion conservation  in many-body theories can indeed be achieved by allowing  time-dependence of the Bogolibov transformation~\cite{PalF}. In the saddle point approximation, however,  
one gets a formulation  close to the quasi-chemical equilibrium theory of superconductors developed by the Sydney group~\cite{Blat}, in which fermion number is explicitly preserved. 
Since the latter approach is in our opinion  more transparent than that of BCS and Bogoliubov-Valatin from a physical point of view,  establishing a connection between superconductivity and superfluidity, we report  a brief account of both methods in Appendix A.

The plan of the work is as follows. In Section 2 we report some considerations about the fermion determinant  with special regard to its form in the presence of stationary gauge fields. In Section 3 we report the definitions and notations we will use and  in Section 4 the time-dependent gauge field-dependent Bogoliubov transformations. In Section 5 we write and solve the saddle point equations, in Section 6 we perform the perturbative expansion in the gauge coupling constant for large chemical potential and we conclude with some  remarks in Section 7.

\section{The sign problem}

In this section we will review the well known {\em sign problem} which affects numerical simulations in presence of fermions in many problems and in the particular case which is interesting for us, namely QCD at finite density. 

We begin from the expression of the grand-canonical partition function of QCD directly in continuous space-time.  
Formally it can be represented as a path integral in euclidean space 
\be
{\mathcal Z} = \int \,[dA]\, \exp(- S_G[A]) {\mathcal Z}_F[A]\, ,
\ee
where $A$ represents the gauge fields, $S_G$ is their pure action, and the fermion partition function is given by a Berezin integral,
\be
{\mathcal Z}_F[A] = \int  \,[d\psi\,  d\bar\psi]\,\exp(- S_F[A])\, ,
\ee
where the fermion action $S_F[A]$ is bilinear in the Grassmann variables $\bar\psi, \psi$. 

The explicit integration on the fermion fields provides the so-called {\em fermion determinant}
\be
{\mathcal Z}_F[A] = \det [\slashed{\nabla}  + m + \mu \gamma_0]\, 
\ee
where $\slashed{\nabla}$ is the contraction of the covariant derivative, which depends on the gauge fields, with the Dirac $\gamma$-matrices, $m$ is the fermion mass and $\mu$ is the chemical potential.

Remark that as
\begin{align}
(\slashed{\nabla}  + m + \mu \gamma_0)^\dagger = & -\slashed{\nabla}  + m + \mu^* \gamma_0 \\
\gamma_5 (\slashed{\nabla}  + m + \mu \gamma_0) \gamma_5 = & -\slashed{\nabla}  + m - \mu \gamma_0
\end{align}
the fermion partition function is necessarily real both when $\mu$ is vanishing or purely imaginary, because under these conditions
\be
{\mathcal Z}_F[A] =  \det \gamma_5 [\slashed{\nabla}  + m + \mu \gamma_0] \gamma_5 = \det [\slashed{\nabla}  + m + \mu \gamma_0]^\dagger = {\mathcal Z}^*_F[A]\, .
\ee
In the Weyl (chiral) representation for $\gamma$-matrices
\be
\gamma_0 = \left(\begin{array}{cc}0 & 1 \\1 & 0\end{array}\right)\, ,
\qquad \vec{\gamma} = \left(\begin{array}{cc}0 & - i \,\vec{\sigma} \\ i \,\vec{\sigma} & 0\end{array}\right)
\ee
where $\vec{\sigma}$ are the Pauli matrices, the relevant matrix takes a partitioned form,
\be
\slashed{\nabla}  + m + \mu \gamma_0 = \left(\begin{array}{cc}m & \nabla_0 +\mu - i \,\vec{\sigma} \cdot \vec{\nabla} \\
\nabla_0 +\mu + i \,\vec{\sigma} \cdot \vec{\nabla} &  m \end{array}\right)  
\, ,
\ee
which is particularly suitable to reduce the evaluation of the determinant to a space of half dimension, indeed,
\be
 {\mathcal Z}_F[A] 
= \det  \left[ m^2   -  ( \nabla_0 +  \mu +  i\, \vec{\sigma}\cdot \vec{\nabla})(\nabla_0 + \mu - i\, \vec{\sigma}\cdot \vec{\nabla}) \right] 
\label{prima}
\, .
\ee
It soon follows that, in the case of $\mu$  vanishing or purely imaginary, not only the determinant is real, but it is also non-negative.  Indeed,
if we set
\be
X = i \nabla_0 + i \mu + \vec{\sigma}\cdot \vec{\nabla}  \, , 
\ee
then
\be
 {\mathcal Z}_F[A] = \det (m^2 + X^\dagger X) \geq \det X^\dagger X \geq 0\, ,
\ee
where the first inequality becomes an equality for vanishing mass and the second whenever $X$ has a vanishing eigenvalue.

But for real, non-vanishing, chemical potential the fermion determinant appears, in general, to be complex. A more detailed analysis can show that it is possible to combine gauge configurations in pairs so that the sum of the determinants be real~\cite{Tomeu}. It remains, however, a possible sign, which
is the problem in dynamic Monte Carlo simulations where fermions are integrated out, because the resulting factor cannot be used as a positive weight which drives the importance sampling of gauge configurations.

\subsection{More on the fermion determinant}

In the Introduction, we have already seen that we are interested in {\em stationary} gauge-field configurations in which the chromoelectric field  vanishes.
In this Subsection we will investigate what are the consequences for the evaluation of the fermion determinant when we restrict ourselves to such gauge configurations, that is such that
\be
[\nabla_0, \vec{\nabla}] = 0\,  \label{condition}
\ee
the spatial covariant derivatives commute with the temporal one.
Within this ensemble of configurations, the fermion determinant becomes
\be
 {\mathcal Z}_F[A] =  \det[\, m^2 - (\vec{\sigma}\cdot\vec{\nabla})^2 - (\nabla_0 + \mu)^2\, ] =
\det[\, H^2 - (\nabla_0 + \mu)^2\, ]\, 
\ee
where the square of the Hamiltonian $H$, which is
\be
H^2 = m^2 - (\vec{\sigma}\cdot\vec{\nabla})^2\, ,
\ee
depends only on the spatial components $\vec{A}$ of the gauge field and is positive definite.

The operator $(\nabla_0 + \mu)$, instead, depends only on the temporal component $A_0$ of the gauge field. It is quite simple to see that if it has, for real $\mu$, a complex eigenvalue $\lambda$, it has also the conjugate one, and therefore the fermion determinant can be written as a product of positive terms
\be
 {\mathcal Z}_F[A] = \prod_{h}  \prod_{\lambda: \Im(\lambda)>0} (h^2 - \lambda^2)(h^2 - \bar{\lambda}^2) >0\, ,
\ee
where $h^2$ are the  eigenvalues of $H^2$.

A few remarks: 
\begin{itemize}
\item the determinant becomes a function of $\mu^2$, that is it does not depend on the sign of $\mu$;
\item the determinant does not feel the sign of $A_0$, so that the average values of the Polyakov lines in positive and negative time directions are the same;
\item in the absence of the chromoelectric field,  the temporal component $A_0$ of the gauge field, which is responsible for the implementation of the Gauss constraint, appears  only in the fermion determinant. The previous discussion implies that
\be
\int [dA_0] \, {\mathcal Z}_F[A]  > 0\, . \label{zpos}
\ee
\end{itemize}

%
%


When using lattice regularization, as we shall do in the following of this paper, strictly speaking, condition~\reff{condition} cannot be used. Covariant derivatives must be replaced by their discretized versions, as for example, $\nabla_\mu^{\pm}$ defined according to
\begin{align}
\nabla_\mu^+ \phi(x) := &  \, U_\mu(x) \phi(x+\hat{\mu}) - \phi(x) \\
\nabla_\mu^- \phi(x) := &  \, \phi(x) - U_\mu^\dagger(x-\hat{\mu}) \phi(x-\hat{\mu}) \, .
\end{align}
The curvature fields are computed by taking the commutators of the covariant derivatives, so that, for example
\begin{align}
[\nabla_\mu^+,\nabla_\nu^+]\, \phi(x) = & \,  [U_\mu(x) U_\nu(x+\hat{\mu}) U_\mu^\dagger(x+\hat{\nu}) U_\nu^\dagger(x) - 1] \\
& U_\nu(x) U_\mu(x+\hat{\nu}) \phi(x+\hat{\mu}+\hat{\nu})\\
= & \, [U_{\mu,\nu}(x) -1]\, U_\nu(x) U_\mu(x+\hat{\nu}) \phi(x+\hat{\mu}+\hat{\nu}) \\
[\nabla_\mu^-,\nabla_\nu^+]\, \phi(x) = & \,  [1 - U_\mu^\dagger(x-\hat{\mu}) U_\nu(x-\hat{\mu}) U_\mu(x-\hat{\mu}+\hat{\nu}) U_\nu^\dagger(x) ]\\
& U_\nu(x) U_\mu^\dagger(x-\hat{\mu}+\hat{\nu}) \phi(x-\hat{\mu}+\hat{\nu})\\
= & \,[1 - U_{-\mu,\nu}(x)]\, U_\nu(x) U_\mu^\dagger(x-\hat{\mu}+\hat{\nu}) \phi(x-\hat{\mu}+\hat{\nu})
\end{align}
where $U_{\pm\mu,\pm\nu}(x)$ are the four plaquettes open at the vertex $x$ in the plane $(\mu, \nu)$. They transform, under a gauge transformation, according to
\be
U_{\pm\mu,\pm\nu}(x) \to \Omega^\dagger(x)\, U_{\pm\mu,\pm\nu}(x) \Omega(x)\, .
\ee
The vanishing of the chromoelectric field means that the spatial link configurations at successive times are related by
\be
U_i(x+\hat{0})  = U_0^\dagger(x) U_i(x) U_0(x+\hat{i})
\ee
so that the open spatial plaquettes, which determine the chromomagnetic fields, are constrained to be gauge equivalent, indeed
\be
U_{\pm i,\pm j}(x+\hat{0}) = U_0^\dagger(x)\, U_{\pm i,\pm j}(x) U_0(x)\, ,
\ee
and therefore their trace is time invariant and they give one and the same contribution to the gauge-field action at all times.

\section{Definitions and notations}

The partition function of lattice QCD can be written as
\be
{\mathcal Z} = \int \,[dU]\, \exp(- S_G[U])\, {\mathcal Z}_F[U]\, ,
\ee
where $[dU]$ is the Haar measure over the Wilson link variables $U$, that lives in the gauge group, $S_G$ is the Wilson action for the gauge fields and ${\mathcal Z}_F[U]$ is the fermion determinant.
For our purposes we shall make use of the operator formulation, so that 
\begin{equation}
\mathcal{Z}_F = {\Tr}^{F}  
\prod_{t=0}^{L_0/2-1}  \, 
{\cal T}_{t,t+1} \,.
\label{part}
\end{equation}
In the above equation $L_0$ is the size extension of the lattice in the temporal direction, so that it is also the inverse temperature, $\cal T$
is the fermion transfer matrix which acts in the Fock space of fermions and  ${\Tr}^F$ is its trace  . We shall make use of the Kogut-Susskind formulation for lattice fermions, so that fermion fields live on blocks of size twice the lattice spacing. The index $t$ labels the blocks along the temporal direction.

The expression of the transfer matrix in the gauge $U_0 = \bone$,  but for Wilson fermions and in the particular case $r=1$ for the Wilson parameter, was given by Lus\"cher~\cite{Lusc}, who proved also its positivity. See also~\cite{Creutz1,Creutz2} for the generalization also to different values of the parameter $r$.  The extention to Kogut-Susskind fermions, in the so-called {\em spin} basis, was given in~\cite{Weisz,SD}. We shall use, instead, the {\em flavour} basis because a simpler transfer matrix is avalaible for this formulation~\cite{PaluF}.

Without fixing the gauge,  the transfer matrix, at nonzero chemical potential $\mu$, can be written as
\be
{\cal T}_{t,t+1} :=  \hat{T}^\dagger_t  \, {\hat V}_t 
\exp(2\,\mu \, \hat{n})\, \hat{T}_{t+1} \,. \label{calt}
\ee
where $\hat{n}$ is the fermion number operator 
\be
\hat{n} :=  \hat{u}^{\dagger} \hat{u} - \hat{v}^{\dagger} \hat{v} \, , \label{hatn}
\ee
(the sum on all the indices is understood) with $\hat{u}^\dagger$ and $\hat{v}^\dagger$, creation 
operators of fermions and antifermions, obeying canonical anti-commutation relations and 
\be
\hat{T}_t  = 
\exp[\hat{v} N_t  \, \hat{u}] \, , \qquad
{\hat V}_t = \exp  [ \hat{u}^{\dagger}\ln  U_{0,t}\, \hat{u} + \hat{v}^{\dagger}  \ln
U_{0,t}^* \, \hat{v} ] \label{transfer}
\,.
\ee
The matrices 
$N_t$ are functions of the  spatial link variables at time $t$. More precisely
\be
N =
 -2  \,  (\gamma_0\otimes \bone)  \left\{   m   +  \sum_{j=1}^3   (\gamma_j   \otimes \bone)  
 \left[  P^{(-)}_j   \nabla_j^{(+)}  + P^{(+)}_j \nabla_j^{(-)}
\right]  \vphantom{\sum_{j=1}^3} \right\} \,  \label{Dirac}
\ee
where
\be
P^{(\pm)}_{j} =  \frac{1}{2} ( \bone \otimes \bone 
\pm \gamma_j \gamma_5 \otimes t_5 t_j ) 
\ee
are projection operators, $\gamma_{\mu}$ and $t_{\mu}$ are Dirac and taste matrices,  
\be
\nabla_j^{(+)} =  \frac{1}{2}  \left( U_j \,T^{(+)}_j  - 1 \right) \quad , \quad
 \nabla_j^{(-)} =  \frac{1}{2}  \left( 1- T^{(-)}_j  U_j^{\dagger}\right) 
\ee
are covariant derivatives, 
$T_j^{(\pm)}$ are forward / backward translation operators of one block  of size twice the lattice spacing and $U_j$ the $j$-th component of $\vec{U}$, the spatial link variables associated to the blocks.


The  operators 
\be
P_{\pm} =  \frac{1}{2} ( \bone \otimes \bone 
\mp \gamma_0 \gamma_5 \otimes t_5 t_0 )
\ee  
project on the  components of the fermion field which propagate forward or backward in time
\begin{eqnarray}
u &=& P_{+} \psi
\nonumber\\
v^{\dagger} &= &P_{-} \psi \, .
\end{eqnarray}
The symbol
 ``$\mbox{tr}$" denotes  the trace over fermion-antifermion internal quantum numbers and spatial coordinates (but not over time).
We introduce the notation, which we will use  for any matrix $\Lambda$
\begin{equation}
\mbox{tr}_{\pm} \Lambda := \mbox{tr}\left( P_{\pm} \Lambda \right) \,.
\end{equation}
Finally we will denote by $T_0^{(\pm)}$  the forward and backward
translation operators of one block, that is two lattice spacing, in the time direction 
\begin{equation}
[T_0^{(\pm)}]_{t_1,t_2 }=\delta_{t_2,t_1 \pm 1} \,. \label{T0}
\end{equation}

 \section{Time-dependent  Bogoliubov transformations}
 
 We evaluate the trace of the fermion transfer matrix in a basis obtained by performing Bogoliubov transformations 
 on the coherent states
\begin{equation}\label{coherent}
| \alpha,\beta\rangle = \exp (- \alpha \, {\hat u}^{\dagger} - \beta \, {\hat v}^{\dagger} ) | 0 \rangle,
\end{equation}
where the $\alpha ,\beta$ are Grassmann fields. 

In a first transformation,  we  introduce quasiparticles operators ${\hat \alpha}, {\hat \beta} $ which have the same fermion number as the original operators $ {\hat u}, {\hat v}$
\be
{\hat \alpha}  =  R^{\frac{1}{2}}\left( {\hat u} -  
 \,{\mathcal F}^{\dagger}  \, {\hat v}^{\dagger}\right) \,, \qquad 
{\hat \beta}  =  \left( {\hat v} +  
{\hat u}^{\dagger} \,{\mathcal F}^{\dagger} \right) \giro{R}{\hspace{-1mm}}^{\frac{1}{2}} \label{Bogoliubov} \\
\ee
where
\begin{equation}
 R= (1 + {\mathcal F}^{\dagger} {\mathcal F})^{-1} \qquad  \giro{R} = (1 + {\mathcal F} {\mathcal F}^{\dagger})^{-1} \, .
\label{involution}
\end{equation}
The upperscript circle denotes the involution defined by the above equations.  
The new operators satisfy 
 canonical commutation relations for any choice of the matrix $ {\mathcal F}$. 
 The vacuum of the new operators is a condensate of this  composite boson
 \be
  |\mathcal{F} \rangle= \exp (\hat{\mathcal{F}}^{\dagger}) \, |0 \rangle \label{defF}
  \ee
where
\begin{equation}
{\hat \fc }^{\dagger} = \uh^\dag \fc^\dag  \vh^\dag \, .
\end{equation}

By a second Bogoliubov transformation, we introduce new quasiparticle operators $ {\hat \sigma}$ which have the same fermion number as ${\hat u}$
\be
{\hat \sigma} = r^{\frac{1}{ 2}} \left( {\hat \alpha}  - {\mathcal D}^{\dagger} {\hat \alpha}^{\dagger}\right)\,,  \label{Bogoliubov2} 
\ee
where
\be 
r= (1+ {\mathcal D}^{\dagger}  {\mathcal D} )^{-1} \,
\ee
and the bosonic field represented by the antisymmetric matrix ${\mathcal D}$  has fermion number two. 
The corresponding operator
\be
\hat{ {\mathcal D}}^{\dagger}  = {\hat \alpha}^{\dagger}{\mathcal D}^{\dagger}  {\hat \alpha}^{\dagger} \label{hatD}
\ee
will represent diquarks.

The vacuum of the new operators is 
\be
|{\mathcal D} , {\mathcal F} \rangle = \exp \left(\frac{1}{ 2} \hat{ {\mathcal D}}^{\dagger}   \right)
\, \exp (\hat{\mathcal{F}}^{\dagger}) \, |0 \rangle = 
\exp \left(\frac{1}{2}  {\hat \alpha}^{\dagger}{\mathcal D}^{\dagger}  {\hat \alpha}^{\dagger}    \right) \exp \left( {\hat u}^{\dagger}{\mathcal F}^{\dagger}  {\hat v}^{\dagger}  \right) |0 \rangle \,,
\ee
namely a condensate of Cooper pairs of quasiparticles in a background field.

If we perform a gauge transformation
\be
\hat{\psi}(x) \to \hat{\psi}^\prime(x) = g(x)\, \hat{\psi}(x)
\ee
both components $\hat{u}$ and $\hat{v}^\dagger$ transform in the same way, that is
\be
\hat{u}(x) \to \hat{u}^\prime(x) = g(x)\, \hat{u}(x)\, , \qquad \hat{v}^\dagger(x) \to \hat{v}^{\prime\dagger}(x) = g(x)\, \hat{v}^\dagger(x)\, .
\ee
In order to get that also $\hat{\alpha}$ transforms in the same way we need, because of~\reff{Bogoliubov}, that the matrix appearing in the first transformation at time $t$ transforms according to
\be
({\mathcal F}_t^\dagger)_{{\bf x}, {\bf y}} \to ({\mathcal F}_t^{\prime\dagger})_{{\bf x}, {\bf y}} = g(t, {\bf x}) \, ({\mathcal F}_t^\dagger)_{{\bf x}, {\bf y}} \, g^\dagger(t, {\bf y}) \, . \label{gif}
\ee
As a consequence, $\hat{\mathcal F}$ and the states $|{\mathcal F}\rangle$ are gauge invariant.

If we now demand that also $\hat{\sigma}$ transforms as $\hat{\alpha}$ we need, because of~\reff{Bogoliubov2}, that the matrix appearing in the second transformation at time $t$ transforms according to
\be
({\mathcal D}_t^\dagger)_{{\bf x}, {\bf y}} \to ({\mathcal D}_t^{\prime\dagger})_{{\bf x}, {\bf y}} = g(t, {\bf x})\, ({\mathcal D}_t^\dagger)_{{\bf x}, {\bf y}} \, g(t, {\bf y}) \, , \label{gid}
\ee
which also implies that $\hat{\mathcal D}$ is gauge invariant.
Therefore under the conditions~\reff{gif} and~\reff{gid} all the states $|{\mathcal D} , {\mathcal F} \rangle$ are gauge invariant.

The trace of the fermion transfer matrix, after the first transformation, was represented in~\cite{CPV, noi-prd} as a Berezin integral, with the result
\begin{eqnarray}\label{Z}
{\mathcal Z}_F &=
\int [d\alpha\,d\alpha^*\,d\beta\,d\beta^*]\, e^{-S_{me} (\fc)-S_{qp}(\alpha,\beta;\fc) }
\end{eqnarray}
where the Grassmann variables $\alpha^*, \alpha, \beta^*, \beta$ satisfy antiperiodic boundary conditions in time.
In the above equation $S_{qp}(\alpha,\beta;\fc)$,  which is the action of quasi-particles,  takes the form 
\begin{multline}\label{SF}
S_{qp}(\alpha,\beta;\fc)=  - 2 \sum_{t=0}^{L_0/2-1}\Big[ \beta_{t+1} {I }_{t+1}^{(2,1)} \alpha_{t+1} + \alpha^*_{t}{I }_t^{(1,2)} \beta^*_{t}\\
	 + \alpha^*_t  (\nabla_t -{\cal H}_t) \alpha_{t+1}-\beta_{t+1}  (\giro{\nabla}_t-\giro{{\cal H}_t)}\beta^*_{t}\Big]  \,
\end{multline}
where the covariant derivatives are defined as
\be
\nabla_t  \, := \frac{1}{2}\,\left(e^{2\mu}\, U_{0,t} -T^{(-)}_0\right) \,, \qquad
\giro{\nabla}_t \, := \frac{1}{2}\,\left( e^{-2\mu}\, U_{0,t}^\dagger-T^{(+)}_0\right)\, ,
\ee
and $T^{(\pm)}_0$ are translation operators of one time block defined in~\reff{T0}. The presence of the factors 2 is related to the fact that neighboring blocks stay at two lattice spacings.

The explicit expressions for the mesonic action $S_{me} (\fc)$, for the Hamiltonians, respectively of  the fermions and the antifermions, ${\cal H}_t$ and $\giro{{\cal H}_t}$, and for the mixing terms between  quasi-particles and quasi-antiparticles ${I }_{t+1}^{(2,1)}$ and ${I }_t^{(1,2)}$
will be reported in the next Section. 

In the present work we are interested in the trace of the fermion transfer matrix after the second transformation. At  this stage, as we will concentrate on the leading contribution in the nilpotency expansion, we restrict ourselves to the bosonic part of the action
\be
S_{bo} = S_{me} + S_{dq}\, ,
\ee
and we disregard the action of the new quasiparticles appearing after the second transformation. More precisely we shall restrict to
\be
\exp  (- S_{bo} ) :=  \, \prod_{t=0}^{L_0/2-1} \frac{\langle {\mathcal D}_t, {\mathcal F}_t | {\mathcal T}_{t,t+1} | {\mathcal D}_{t+1}, {\mathcal F}_{t+1} \rangle }{\langle {\mathcal D}_t, {\mathcal F}_t | {\mathcal D}_t, {\mathcal F}_t \rangle } \, .
\ee
$S_{me}$, which can be obtained by putting ${\mathcal D}_t = 0$ at all times $t$, is what we got previously, and the extra term $S_{dq}$, that we call the diquark action, is derived in Appendix C, and takes the form
\be
S_{dq} =  \frac{1}{2} \sum_{t=0}^{L_0/2-1} \mbox{tr}_+ \left\{
  \ln \left( 1+ {\mathcal D}_t {\mathcal D}_t^{\dagger} \right) 
  - \ln \left( 1+ e^{4\,\mu}\,{\mathcal D}_t  \,  Q_{t+1,t}^{-1} \, {\mathcal D}^{\dagger}_{t+1}  \, {Q}_{t+1,t}^{-T}\right)  \right\}   
\ee
where the matrix $Q_{t+1,t}^{-1}$ is defined by
\be
Q_{t+1,t}^{-1} :=  U_{0,t} - 2 \, e^{-2 \mu}\,{\mathcal H}_t \, , \label{defQ}
\ee
and we denote by ${Q}_{t+1,t}^{-T}$ the transpose of the inverse of the matrix ${Q}_{t+1,t}$ (which is also the inverse of the transpose).

In the following we will work in the gauge $U_{0,t}=1$, for all times $t$, and we should impose the Gauss constraint on the states $| {\mathcal D}_t, {\mathcal F}_t \rangle $. But  we don't need it, because, as a consequence of their gauge invariance, they are left invariant by the Gauss projector ${\mathcal P}_G$, which takes into account also the transformations~\reff{gif} and~\reff{gid}, that is
\be
{\mathcal P}_G \, | {\mathcal D}_t, {\mathcal F}_t \rangle  = | {\mathcal D}_t, {\mathcal F}_t \rangle \, 
\ee
for all times $t$.

It is important to call the attention of the reader on a relevant difference between the conditions~\reff{gif} and~\reff{gid} for the gauge invariance of, respectively, $\hat{\cal F}$ and $\hat{\cal D}$. Indeed, in order to ensure the gauge invariance of the operator $\hat {\cal F}$, which acts on the fermion Fock space at time $t$, it is enough to allow a dependence of ${\cal F}$ from the spatial-link variables $U_{{\mathbf k}, t}$, such that, under a gauge transformation
\be
{\mathcal F}_t^\dagger[U_{{\mathbf k}, t}] \to {\mathcal F}_t^{\prime\dagger}[U_{{\mathbf k}, t}^\prime] = {\mathcal F}_t^{\dagger}[U_{{\mathbf k}, t}^\prime]
\ee
with
\be
({\mathcal F}_t^{\dagger}[U_{{\mathbf k}, t}^\prime])_{{\bf x}, {\bf y}} = ({\mathcal F}_t^{\dagger}[g_t\, U_{{\mathbf k}, t}\, g_t^\dagger])_{{\bf x}, {\bf y}} 
= g(t, {\bf x}) \, ({\mathcal F}_t^\dagger[U_{{\mathbf k}, t}])_{{\bf x}, {\bf y}} \, g^\dagger(t, {\bf y})\, .
\ee
This indicates that in the matrix element  the dependence on the link variables is  by strings $\Gamma_t({\cal C_{{\bf x}, {\bf y}}})$  of products of link variables at time $t$ along  paths ${\cal C}_{{\bf x}, {\bf y}}$ between the positions $\bf x$ and $\bf y$, which realize the {\em parallel transport} and transform according to
\be
\Gamma_t({\cal C_{{\bf x}, {\bf y}}}) \to g(t, {\bf x}) \, \Gamma_t({\cal C_{{\bf x}, {\bf y}}})\, g^\dagger(t, {\bf y})\, .
\ee

An attempt to proceed in the same way with the operator $\hat{\cal D}$ does not work. For example, for the case in which the gauge group is $SU(3)$, in the simple proposal 
\be
({\mathcal D}_t^\dagger)_{{\bf x}, {\bf y}}^{a, b} =  \epsilon_{a'b'c}\, \left(\Gamma_t^{a'a}({\cal C}_{{\bf w},{\bf x}})\Gamma_t^{b'b} ({\cal C}_{{\bf w}, {\bf y}})
-\Gamma_t^{a'b}({\cal C}_{{\bf w}, {\bf y}}) \Gamma_t^{b'a}({\cal C}_{{\bf w}, {\bf x}}) \right) \, d^c(t, {\bf w}) \, ,
\ee
where $\cal D$ is explicitly antisymmetric, as it must be, and we have also indicated the gauge-group indices, the gauge-invariance condition~\reff{gid} forces $d^c(t, {\bf w})$ to transform, under gauge transformations, as a quark field. Therefore, in order to implement the gauge-invariance of $\hat{\cal D}$ we must introduce a new dynamical field for diquarks, which acts as a {\em compensating} field, according to the general discussion on symmetry breaking we presented in~\cite{noi-prd}.

\subsection{Expressions for the mesonic and quasi-particles action}

By using the matrix ${\cal F}$, which defines the first Bogoliubov transformation, and the matrix $N$, which appears in the definition of the transfer matrix~\reff{transfer}, we introduce the shorthands
\be  
\NdB := 1+N^{\dag}_{t} \fc_{t}\,, \label{FN}   \qquad
 \BNd \,  :=  \, 1+\fc_{t} N^{\dag}_{t}
\ee
in terms of which we define the expressions
\begin{align}
E_{t+1,t} \, := \, & {\cal F}_{N,t+1}^\dagger U_{0,t}^\dagger {\cal F}_{N,t} + {\cal F}_{t+1}^\dagger U_{0,t}^\dagger {\cal F}_{t}  \label{E} \\
\giro{E}_{t+1,t} \, := \, & \giro{{\cal F}} _{N,t} U_{0,t} \left(\giro{{\cal F}} _{N,t+1}\right)^\dagger + \giro{{\cal F}}_{t} U_{0,t} \left( \giro{{\cal F}}_{t+1}\right)^\dagger \, .
\end{align}
With the help of the definitions of  $\giro{R}$ and $R$, given in~\reff{involution}, we report now the expression 
for the Hamiltonians for the fermions and the antifermions
\begin{align}
{\cal H}_t \,:= &\, \frac{1}{2}\,e^{2\mu}\left( U_{0,t} - \mt^{-\frac{1}{2}}\me_{t+1,t}^{-1}\mc_{t+1}^{-\frac{1}{2}} \right) \label{hstorto}\\
\giro{{\cal H}}_t\, :=& \,  \frac{1}{2}\,e^{-2\mu}\left( U_{0,t}^\dagger - \nc\hspace{-1mm}{}^{-\frac{1}{2}}_{t+1} \giro{\me}\hspace{-1mm}{}_{t+1,t}^{-1}\nt^{-\frac{1}{2}} 
\right)  \, ,
\end{align} 
and for the mesonic action
\be
\label{bosonaction}
   S_{me}(\fc):= - \, \sum_{t=0}^{L_0/2-1}\tr_{+} \ln\left ( R_t \, \me_{t+1,t}  \right) = -  \, \sum_{t=0}^{L_0/2-1}\tr_{+} \ln Q_{t+1,t}
\ee
where we used the definition~\reff{defQ}, and for the the terms  which mix  quasi-particles with quasi-antiparticles 
\begin{align} 
  \Iaa_t^{(2,1)} & := 
 \, \frac{1}{2}\,  \nt^{\frac{1}{2}}\,  \left[ \nt-\giro{\me}\hspace{-1mm}{}_{t,t-1}^{-1}\BNdtm  U_{0,t-1}\right] \fc_{t}^{\dag -1}  \mc_{t}^{\frac{1}{2}} \\
 \Iaa_t^{(1,2)} &:= 
  \, \frac{1}{2}\, \mc_{t}^{\frac{1}{2}}\,  \fc_{t}^{-1}\left[ \nt- U_{0,t}\left(\BNdt\right)^\dag \giro{\me}\hspace{-1mm}{}_{t+1,t}^{-1}\right]\, \nt^{\frac{1}{2}} \, .
\end{align}
As a consequence of the definitions~\reff{defQ} and~\reff{hstorto} we get the relation
\be
Q_{t,t+1} := R_{t+1}^\frac{1}{2} E_{t+1,t} R_t^\frac{1}{2}\, .
\label{newdefQ}
\ee

\section{Saddle point equations} 

We determine the values of the background and diquark fields by minimizing the bosonic part of the effective action.
The stationarity equations are
\be
\frac{\partial }{\partial {\mathcal F}_t} \,  S_{bo} = \frac{\partial }{\partial {\mathcal D}_t} \, S_{bo}  = 0 \, , \label{saddle}
\ee
We remark that in the nilpotency expansion {\em we have to integrate, on an arbitrary measure, both on the background and the diquark fields}.
Then stationarity equations become saddle point equations in the nilpotency expansion. There is a subtlety. In the latter case we should also add the equations
\be
\frac{\partial }{\partial {\mathcal F}_t^{\dagger}} \,  S_{bo} = \frac{\partial }{\partial {\mathcal D}_t^{\dagger}} \, S_{bo}  = 0 \, ,
\ee
and consider the fields and their Hermitian conjugates independent in the variations. 
It will turn out that the solutions are Hermitian fields, so it will be enough to consider the solutions of~\reff{saddle} by keeping fixed the Hermitian conjugate fields.

 
\subsection{In the absence of the diquark field}

In the absence of the diquark field, that is whenever ${\cal D}_t=0$ for every time $t$, the background field has already been determined~\cite{noi-prd}. We outline its derivation. 
We look for solutions stationary in time, as appropriate to the vacuum. If $\cal F$ is stationary, the elementary bosonic fields coupled to the fermions which enter its expression should also be stationary. In gauge theories $\cal F$  must certainly depend on spatial link variables $U_{\bf k} (t, {\bf x})$. Stationarity in time for gauge fields can be formulated in a gauge covariant way by requiring that these fields evolve according to gauge transformations, that is
\be
U_{\bf k} (t, {\bf x}) = W^{\dagger}_{t, {\bf x}} U_{\bf k }(0, {\bf x})W_{t, {\bf x}+{\hat {\bf k}}}  \label{spatial} \, .
\ee
As a consequence, the chromomagnetic contribution to the pure gauge-field action, namely, the trace of spatial plaquettes, does not depend on time.

Accordingly, the matrices ${\mathcal F}_t$ and $N_t$ are related to that at time $t=0$, that is, if ${\mathcal F}_0 = {\mathcal F}$ and $N_0 = N$ then
\be 
{\mathcal F}_t = W_t^\dagger {\mathcal F} W_t \, , \qquad N_ t = W_t^\dagger N W_t \, .
\ee
We still wish to set the contribution of the chromoelectric field to the gauge-field action, namely, the trace of spatio-temporal plaquettes, to be independent on time. We have been able to arrive at a stationary solution for $\cal F$ only with the particular choice
 \be
 W_{t+1, {\bf x}} = U_0(0, {\bf x}) U_0(1,{\bf x })  \dots U_0(t,{\bf x})  \label{simpl} \, 
\ee
which lets the contribution from the chromoelectric field vanish at all times. 

Under these conditions the saddle point equations for the background field become independent of time
\be
{\mathcal F} =  N + {\mathcal F} \,\left( {\mathcal F}_{N}\right)^{-1} \, .
\ee
%
%
The relevant extremal solution for the background field is
\begin{equation}
\overline{{\mathcal F}}= N(2N^{\dagger}N)^{-1}  \left[ N^{\dagger}N
  + \sqrt{ (N^{\dagger}N)^2 + 4  N^{\dagger}N} \right]  \label{Fsaddle} \, .
\end{equation}
This is also the solution of the equations~\cite{CPV}  
\be
 		\Iaa{}_t^{(2,1)}\,=\,\Iaa{}_t^{(1,2)}=0\,.
\ee
This means that at the minimum of the vacuum energy there is no quasiparticle-antiquasiparticle mixing. (This is in close analogy to  the case  of the Bogoliubov transformation in the BCS theory, as explained in Appendix A.2. But needless to say, unlike the latter these terms do not violate any symmetry). 
Hence at the saddle point, the effect of the Bogoliubov transformations~\reff{Bogoliubov} is analogous to that of the Foldy-Wouthuysen transformations which separate positive from negative energy states in the Dirac Hamiltonian~\cite{CPV}.

The time evolution of the quasiparticle Hamiltonians is slightly different:
\be
{\mathcal H}_t =   W_t^\dagger {\mathcal H} W_{t+1}\, , \qquad \giro{{\mathcal H}}_t = W_{t+1}^\dagger \giro{{\mathcal H}} W_t \, 
\ee
and similarly
\be
Q_{t+1,t}^{-1} = W_t^{\dagger} \, Q^{-1} \, W_{t+1}\, .         \label{Qstatio}
\ee
At the saddle point, the quasiparticle and antiquasiparticle Hamiltonians at time $t=0$, respectively ${\overline {\mathcal H}}$ and $\giro{{\overline {\mathcal H}}}$, are simply related by
\be
2\, e^{-2 \mu}\, {\overline {\mathcal H}} =  2\,e^{2 \mu} \giro{{\overline {\mathcal H}}} =  1 - \overline{Q}^{-1} = 1- {\overline {\mathcal F}}_N^{-1} \, .
\label{Q1}
\ee
They are  Hermitian functions of  $N^{\dagger} N$, and the vacuum energy is
\be
{\overline S}_{me} = - \frac{L_0}{2}  \, \mbox{tr}_+ \ln  \overline{Q} \,.
\ee
%

\subsection{In the presence of the diquark field}

Now we rewrite the diquark field action exploiting the time dependence~(\ref{spatial}) of the spatial-link variables, but we will not need the explicit expression of the background field. So we will write, in addition to~\reff{Qstatio} where $Q = Q_{1,0}$,
%
\be
{\mathcal D}_t= { W}_t^T \, {\mathcal D} \, W_t
\ee
where ${\mathcal D} = {\mathcal D}_{0}$. Then the  time dependence disappears from the bosonic  action 
\be
S_{bo} =  \frac{L_0}{2}  \,  \mbox{tr}_+ \left\{- \ln  Q 
+ \frac{1}{ 2} \ln \left( 1+ {\mathcal D}^{\dagger} {\mathcal D} \right) \right.
 - \left. \frac{1}{ 2} \ln \left( 1+ e^{4 \mu}\,{\mathcal D}  \,  Q^{-1} \, {\mathcal D}^{\dagger}  \, { Q}^{-T}\right)  \right\} \,.
\ee 
We first remark that all the dependence on the background field $\cal F$ is now contained in the dependence from $Q$, so that
\be
\frac{\partial S_{bo}}{\partial {\mathcal F}} = \frac{\partial S_{bo}}{\partial Q}\frac{\partial Q}{\partial {\mathcal F}} = 0\, .
\ee
As the equation
\be
\frac{\partial Q}{\partial {\mathcal F}} = 0
\ee
does not involve the diquark field $\cal D$ and its relevant solution for $\cal F$ is exactly the $\overline{\mathcal F}$ given in~\reff{Fsaddle}, the background field does not depend on the diquark one.

The stationarity equation for the diquark field is
\be
{\mathcal D} = e^{4 \mu}\, { Q}^{-T} {\mathcal D} \, Q^{-1} \label{Dsella}
\,.
\ee
As the bosonic action is gauge invariant, but the field ${\cal D}$ is not, the stationarity equation determines only the class of ${\cal D}$ equivalent under gauge transformations.

In order to analyze the solutions of~\reff{Dsella}
we construct the diquark structure function in the basis of  eigenstates of the quasiparticle Hamiltonian for given gauge field configurations, which according to~\reff{Q1} are also eigenstates of $\overline{Q}$
\be
\overline{Q} |i \rangle  = Q_i |i \rangle \,.
\ee
Remark that, although the operator $Q$ is nonlocal, its eigenvalues are simply related to those of the local operator $N$, as shown in~\reff{qi}. This explicit relation shows also that they are all real and greater than unity.

The saddle point equations then become 
\be
{\mathcal D}_{ij} = e^{4 \mu}\, Q^{-1}_i {\mathcal D}_{ij}   \, Q^{-1}_j  \,.
\ee
First we notice that this equation can only determine $|{\mathcal D}_{ij}|$ because any possible phase factor cancels from both sides and we can restrict to the case in which ${\mathcal D}_{ij}$ is a nonnegative real number. Apart from the trivial solution ${\mathcal D}_{ij}=0$,  the saddle point equations are satisfied for arbitrary ${\mathcal D}_{ij}$  provided  $Q_i= e^{4 \mu}\,Q_j^{-1}$, in which case the contribution to the action vanishes. The relevant minima of the action are then reached on the boundary of the range of the
$|{\mathcal D}_{ij}|$, namely when these are zero or infinity. In the first case the Bogoliubov transformationtion is the identical transformation,
in the second case it interchanges creation with annihilation operators. 

 So we must determine in which way to chose between zero and infinity for any given eigenstate of $Q$.
 We will not search for the most general form of the ${\mathcal D}$. Instead we restrict ourselves to the following canonical form
 \be
{\mathcal D}_{ij} =  \delta_{j, p(i)}  \, \epsilon_{ij}\, {\mathcal D}_{i} \,,  \qquad  {\mathcal D}_{i} =  {\mathcal D}_{p(i)} 
\ee
 in which any quasiparticle state $i$  is associated to one and only one conjugate state $p(i)$. 
In the many-body language this is called simple pairing. 
Notice that, since the ${\mathcal D}_{ij}$ are antisymmetric, 
they can always be cast into this form, but at the price of a unitary transformation which would change the form of the quasiparticle hamiltonian.  For the same reason, we 
observe  that, since  the ${\mathcal D}_i$ are functions of the spatial link variables,  pairing can be simple only in a given gauge.

 We will denote by $P$ the set of the states $i$ for which $ |{\mathcal D}_{i}| =\infty$. 
 It follows that if $i\in P$ also $p(i) \in P$. Therefore
\begin{eqnarray}
S_{bo} & = & \frac{L_0}{2} \, \left\{  -   \sum_i   \ln Q_i  
+    \frac{1}{ 2} \sum_{i \in P} \left[\ln |{\mathcal D}_{i}|^2
-  \ln \left(  e^{4 \mu}\, |{\mathcal D}_{i}|^2  Q_{i}^{-1} Q_{p(i)}^{-1}\right)\right] \right\}
\nonumber\\
 & = & \frac{L_0}{2}  \, \left\{  -  \sum_i  \ln  Q_i 
-  \sum_{i \in P} \left[ 2\mu - \ln  Q_{i}  \right]  \right\} \label{S_{bo}}
\end{eqnarray}
assuming $ Q_{i}  = Q_{p(i)}$, at least when $i\in P$. For given chemical potential this action is minimal if  $Q_{i} < e^{2\mu} $ for each state $i\in P$. 
The contribution originating from diquarks in the states $i\in P$ cancels the contribution to the action $S_{bo}$ of the corresponding components of the background field. This has a simple physical interpretation: because of Pauli blocking the states occupied by diquarks are not accessible to fermions in the background field. 
At $\mu=0$ as $\overline{Q} >1$ the set $P$ is empty, so that ${\cal D}_i = 0$ for any $i$.  Increasing $\mu$,  the number of fermionic states in the background field is progressively depleted, until a deconfining transition will take place.
 
 
 We can perhaps understand better this result if we rewrite the bosonic energy in the form
\be
S_{bo} =  \frac{L_0}{2}  \, \left\{    \sum_i  \log (1 - 2 e^{- 2 \mu}{\overline {\mathcal H}}_i)  -   \sum_{i \in P} \ln (e^{2 \mu} - 2 {\overline {\mathcal H}_i} )   \right\}\,. 
\ee
Taking the   formal continuum limit we get that $\overline{{\mathcal H}}_i < \mu$ for the states $i\in P$ and
 \be
 {\mathcal Z}_F \approx  e^{ \frac{1}{T} \left( \sum_i \overline{{\mathcal H}}_i   - \sum_{i \in P} ({\overline {\mathcal H}}_i - \mu ) \right)}   \, .
 \ee

We remind that the eigenvalues ${\overline {\mathcal H}}_i  $ are functions of the gauge fields. Integrating over the gauge fields with the
pure gauge fields weight will smooth out the distribution of the values of  $|{\mathcal D}_i| = 0, \infty$.

\subsection{Variational character  of the saddle point approximation}                          

It is important to notice that the expression of the bosonic action can be obtained by evaluating the partition function in a Fock space which contains only the state $|\overline{{\mathcal D}}, \overline{{\mathcal F}} \rangle$. 
At zero baryon density,  ${\cal D}=0$, a variational calculation was performed in~\cite{CLP} and, subsequently, it was recognised to provide the vacuum contribution obtained by a suitable Bogoliubov transformation~\cite{Palu1}.  Here we sketch the proof of this property in the more general case.

Let us start from the expression~\reff{part} for the fermion determinant and choose the gauge $U_{0,t}=1$ for all times $t$, imposing the Gauss constraint on the states by means of the Gauss projector ${\mathcal P}_G$.
%
Under stationarity conditions for the gauge-field configurations we get 
\be
\int [dU_0]\, {\mathcal Z}_F=  \, 
{\Tr}^{F} \,  {\mathcal P}_G\,   \left({\cal T}^\frac{L_0}{2}\right)_{U_{0,t}=1} 
\ee
where we don't need to remember the time indices for the transfer matrix.

Our transfer matrix $\cal T$ at $U_{0,t}=1$ is positive definite, indeed ${\cal T} \sim \hat{T}^\dagger \hat T$, and we simply get the inequality~\reff{zpos} on the lattice
\be
\int [dU_0]\, {\mathcal Z}_F > 0 \, .
\ee
Thanks to the positivity of the transfer matrix we shall now use the inequality 
\be
 {\Tr}^{F} \,  {\mathcal P}_G\,   \left({\cal T}^\frac{L_0}{2}\right)_{U_{0,t}=1}  \geq \left[ \frac{\langle {\cal A} |   \left({\cal T}\right)_{U_{0,t}=1} | {\cal A}\rangle}{\langle {\cal A} | {\cal A} \rangle}\right]^\frac{L_0}{2}
\ee
valid for any gauge invariant state $|{\cal A}\rangle$ in the fermion Fock space. By choosing $|{\cal A}\rangle =  | \overline{{\mathcal D}}, \overline{{\mathcal F}}\rangle$ we get
\be
\int [dU_0]\, {\mathcal Z}_F \geq \,  \left[ \frac{\langle \overline{{\mathcal D}}, \overline{{\mathcal F}} | \left({\cal T}\right)_{U_{0,t}=1} | \overline{{\mathcal D}}, \overline{{\mathcal F}} \rangle }{\langle \overline{{\mathcal D}}, \overline{{\mathcal F}} | \overline{{\mathcal D}}, \overline{{\mathcal F}}\rangle }\right]^{\frac{L_0}{2}} 
= \, \exp  \left[- S_{bo}(  \overline{{\mathcal D}}, \overline{{\mathcal F}}) \right]  \geq \, \exp  \left[- S_{me}( \overline{{\mathcal F}}) \right] \, 
\ee
which shows that the fermion determinant is approximated from below by the exponential of minus the bosonic action $S_{bo}$, which is always greater than the exponential of minus the mesonic action, obtained by putting ${\cal D}=0$ in $S_{bo}$.

%

\section{Perturbative expansion in the gauge coupling constant}

At sufficiently high baryon density an expansion in the gauge coupling constant can be justified. 
We proceed as if a nonvanishing background field will survive. The study of the densities at which the background field vanishes is deferred to a future work. This Section can then be   considered propedeutic to such future work. But it will also give us the possibility of a critical discussion of perturbative calculations in the gauge coupling constant  and of a comparison with the non relativistic case. 

Under our  assumptions
the diquark effective action does not depend on the temporal links, but it depends on the spatial ones through the matrix $Q$  and  the diquark structure functions $ {\mathcal D}$.  In order to determine $ {\mathcal D}$ variationally, we expand $Q$ in powers of the  
the gauge coupling constant
\be
Q^{-1} \approx 1+ A+ g \,B + g^2 \, C\,.
\ee
Setting 
\be
\rho = \frac{ {\mathcal D}^{\dagger} {\mathcal D}  }{  1+ {\mathcal D}^{\dagger} {\mathcal D} }\,, \qquad\psi= 
\, {\mathcal D} \,\frac {1 }{  1+ {\mathcal D}^{\dagger} {\mathcal D} } 
\ee
and {\it expanding the $\ln$ with respect to lattice spacing and gauge coupling constant }we get
\be
S_{bo} \approx  - \, \frac{ L_0}{2}  \, \mbox{tr} \left\{  \ln Q  +  \rho \,(A+ g \,B + g^2 \, C)
   - \frac{1}{2} \,  g^2 \rho \, B \rho \, B  + \frac{1}{2} \, g^2 \psi B \psi^{\dagger} \,{\tilde B} \right\}\,.
\ee
Notice that $\psi$ is antisymmetric. 
The above expression is essentially identical to that of the many-body theory reported in Appendix A, Eq.(\ref{vac energy}), in which the first term is the vacuum energy (independent on the
chemical potential), the second 
the kinetic energy, the third the density-density interaction and the last one the interaction between fermions in one and 
the same Cooper pair. Such {\em close correspondence} exists because we constructed the diquark field in terms of quasiparticles.

At this point   we assume  simple pairing, so that
\be
\left({\mathcal D}^{\dagger}{\mathcal D}\right)_{ij} = \delta_{ij} |{\mathcal D}_i|^2
\, , \qquad \rho_{ij}= \delta_{ij} \rho_i\,, \qquad
(\psi)_{ij}= \delta_{j, p(i)} \, \epsilon_{ij}  \,  
\psi_i
\ee
where
\be
\rho_i = \frac{|{\mathcal D}_i|^2} {1+ | {\mathcal D}_i |^2}\,, \qquad 
\psi_i = \, 
\frac{1}{1+ | {\mathcal D}_i |^2} \, {\mathcal D}_i \,.
\ee
We remind that we use the label $i$ for all quasiparticle labels, that is the position vector $\bf x$, the  Dirac  $\alpha$, the favour $f$ and the colour $a$ indices.
As a concrete example of simple pairing  for the case of 2 flavors we can assume
\be
{\mathcal D}_{ {\bf x},\alpha, f_1 , a, {\bf y},\beta f_2,b} = \epsilon_{f_1f_2} \, \epsilon_{3 a b} \,\epsilon_{\alpha \beta} \,  \delta_{ {\bf x}, {\bf y}}{\mathcal D}_{ {\bf x}}
\ee
\be
B_{ {\bf x}, \alpha, f_1, a, {\bf y}, \beta, f_2,b} = \sum_{k,I} ( \Lambda_I)_{ab} \,( B_k )_{ {\bf x},\alpha, f_1,  {\bf y}, \beta, f_2} (A_k^I)_{{\bf y}}
\ee
\be
\mbox{similarly for }C
\ee
where $k$ labels the spatial components and the $\Lambda_I$ are the gauge group generators.

The effective action now becomes
\begin{multline}
S_{bo} \approx  - \, \frac{L_0}{2}  \,\sum_i\left\{ \vphantom{\frac{1}{2}} \ln Q_i +  \rho_i \,(A+ g \,B + g^2 \, C)_{ii}  - \frac{1}{2} \,  g^2 \rho_i \, B_{ij} \rho_j B_{ji}  \right. \\
\left.    +
 \frac{1}{2} \left(  \psi_i^* \Delta_i + \Delta_i^*  \psi_i \right) \right\}
\end{multline}
where
\be
\Delta_i= \frac{1}{2} \, g^2 \epsilon_{i, p(i)}  \sum_k \epsilon_{k, p(k)} \, B_{ik} \, B_{p(i) p(k)} \, \psi_k  \label{gap}
\ee
is the celebrated gap function.
Usually  in many-body problems the density-density interaction is small but complicates the variational equation, and for this reason it is accounted for renormalizing  phenomenologically the chemical potential~\cite{Blat}. We will adopt this criterion. Then we will integrate over the gauge fields with the pure gauge
measure, and we will denote by $\langle...\rangle$ such average. Note that in the expansion we have disregarded the dependence of the
diquark structure functions on the gauge fields. This can be justified only if the diquark is {\em approximately pointlike}, namely if its mean square radius is much smaller that the average interquark distance (see ref.~\cite{Bailin}). We will say a little more about this in the last Section. Bearing in mind that $\langle B \rangle=0 $ because $B$ is linear in the 
gauge fields, we define 
\be
K_i = \mu_{\rm{eff}} + ( A +  g^2 \langle B^2 \rangle )_{ii} - \frac{1}{2} \sum_k \, g^2  \rho_k  \langle B_{ik}  \, B_{ki} \rangle \,.
\ee
The averaged effective action is
\be
\langle S_{bo} \rangle \approx - \frac{L_0}{2}  \,\sum_i  \left\{  \ln Q_i  +  \rho_i \,(K_i- \mu_{\rm{eff}} )
 +  \frac{1}{2}  \left(\langle  \Delta_i^*\rangle  \psi_i + \psi_i^* \langle \Delta_i \rangle \right)  \right\}\,.
\ee
 Variation with respect to ${\mathcal D}$ (omitting the symbol of average on $\Delta$)gives
 \be
2 (K_i - \mu_{\rm{eff}}) {\mathcal D}_i + \frac{1}{2} \left( {\mathcal D}_i^2 -1 \right) \Delta_i=0
\ee
which has the solutions
\be
\psi_i = \pm \frac{ \Delta_i} {2 \sqrt{ (K_i - \mu)^2 +|\Delta_i|^2 }} \,.
\ee
Inserting them  into the definition of the gap function (\ref{gap}) we get the gap equation
\be
\Delta_i = \pm \frac{1}{2} \,g^2\,\sum_k \frac{ \Delta_k} { \sqrt{ (K_k - \mu)^2  + | \Delta_k|^2 }} 
\,  \epsilon_{i, p(i)}  \, \epsilon_{k, p(k)} \,\langle  B_{ik} \, B_{p(i) p(k)}\rangle \,.
\ee
This expression has two specific features which agree  with the standard result~\cite{Barr,Bailin,Alford,Rapp}. The first is that in both cases the dominant
contribution to the superconducting gap comes from chromomagnetic fields. The second is that these fields are quasistatic in the standard approach, while in the present one they are completely time independent.

\section{Summary and future perspective}

We have investigated QCD  guided by the  theoretical indications that two quarks correlations are important at all baryon densities. We introduced such correlated pairs by means of     Bogoliubov transformations in the formalism of the transfer matrix with lattice regularization. We performed a first transformation which produces a background field and quasiparticles with the quark quantum numbers. A second transformation yields the diquark field in terms of quasiparticles. Evaluating the trace in the partition function using a fermionic basis obtained by these Bogoliubov transformations on fermionic coherent states we got an effective action of the system exactly equivalent to the original one.  At variance with previous use of the Bogoliubov transformations  we let them to depend on time and on spatial-link variables. This makes it possible to enforce for quasiparticles the same symmetry transformations as for quarks, and thus  perform the saddle point approximation keeping gauge invariance manifest. The construction of the diquark field in terms of quasiparticles constitutes another innovation.

We have investigated  the effective theory  at the zeroth order of a nilpotency expansion, namely an expansion in the inverse of the number of fermionic states in the structure functions of the composites, called the index of nilpotency. This means that we derived the effective action in a saddle point approximation, that we have shown to be equivalent to a variational calculation, in which the free energy is minimised with respect to  background and  diquark fields. 

The solution for the background field in the absence of diquarks was found in previous papers. According to this solution the QCD vacuum is a dual superconductor in which the chromoelectric field is totally expelled from the vacuum and the fermion Fock space contains  quasiparticles only in the form of point-like  color singlets~\cite{noi-prd}.

In the present work we have solved the saddle point equations in the presence of a diquark field. In order to describe multiquark mesons and baryons as bound states of a quark and a diquark we should include the fluctuations of this field. We have restricted ourselves, however, to the study of diquark condensation at finite chemical potential. At fixed stationary gauge-field configuration, we derived an expression for the diquark contribution to the free energy that cannot be evaluated analitically  but has a definite positive sign.

Integrating on the space of stationary gauge-field configurations the effective action in the saddle point approximation, by means of an expansion in powers of the gauge coupling constant, a gap equation is obtained compatible with previuos results. The gap is dominated by {\em static chromomagnetic fields}. 

We think, however, that a perturbative expansion in the gauge coupling constant cannot be fully justified even for large values of the chemical potential. Schematically, at the baryon density at which condensation of molecular diquarks is expected, the gauge coupling constant is presumably too large. For baryon densities for which the expansion might be justified, on the other hand, we expect a BCS ground state, in which dibaryons have a size much bigger than the interquark separations, so that the dependence of the diquark structure functions on the spatial gauge fields cannot be ignored.

We hope that our formulation should give a reasonable approximation to the QCD partition function for values of chemical potential of the order of the nucleon mass. Increasing the chemical potential we should meet chiral symmetry restoration, which according to our conjecture should be
accompanied by the vanishing of the background field. If the phase transition is of first order, to determine its location we should compare the free energy evaluated in the present paper with that of the chirally symmetric phase in which the background field should vanish. But we notice that the latter cannot be simply obtained by setting the background field to zero in our equations. In fact in our saddle point approximation we disregarded quasiparticles, appearing after the second Bogoliubov transformation, on the usual, reasonable assumption that they are separated from  the vacuum by a large gap. If instead the background field vanishes, the particle antiparticle mixing in the action is again active, and if we construct the diquark field in terms of particles, we have no reason to expect that a gap exists for antiparticles. We must therefore proceed in a different way that we hope to illustrate in a future work.

\section*{Acknowledgements}

 F. Palumbo has been partially supported by EEC  contract WP22 ``Hadron Physics 2, Study of strongly interacting matter", Grant Agreement n. 227431.
 
We thank Giovanni Viola for his collaboration in a preliminary investigation on the possibility of introducing diquarks in the nilpotency expansion.


\appendix

\section{Superconductivity in many-body systems}

The phenomenon of superconductivity  is explained in terms of Bose-Einstein condensation of fermion pairs,
called Cooper pairs. The first suggestion in this direction was made by Ogg~\cite{Ogg}, who observed persistent ring currents in
 solutions of alkali metals in liquid ammonia. But the   importance his work was not understood and it
did not have any influence in the development of the theory of superconductivity. 

The approach based on this idea is due to the Sidney group~\cite{Blat}, and it is formulated
in the framework of the so called quasi-chemical equilibrium theory, in which there an equilibrium between formation and dissociation of
Cooper pairs. The formalism respects fermion-number conservation, but deals only with the ground state energy. 
In spite of its conceptual simplicity, calculations in this formalism are quite complicated, and  were completed only after the BCS
theory~\cite{Bard} was published. We will also report this theory in the version of Bogoliubov and Valatin~\cite{Bogo,Valatin} , which does not respect fermion conservation, but introduces quasiparticles and  is close to our approach from a technical point of view.

\subsection{Quasi-chemical equilibrium theory}

The quasichemical-equilibrium theory is based on a variational calculation with a wave function $\Psi_N$ constructed in terms 
of the wave function $\varphi[i,j]$ of the pair of the $i$-th and $j$-th  among $N$ fermions
\be
\Psi_N = \frac{C_N}{2^N\, N!} \sum_{\sigma\in {\cal S}_N} \, \sgn(\sigma)\, \varphi[\pi(1),\pi(2)] \dots \varphi[\pi(N-1),\pi(N)] = C_N (\pf \varphi)
\ee
$C_N$ and a normalization constant, ${\cal S}_N$ is the symmetric group of order $N$ and $\sgn(\sigma)$ is the sign of the permutation $\sigma$.  The fermion Hamiltonian is
\be
H_N= \sum_i \frac{ {\bf p}^2_i}{2M} + \sum_{i<j} V_{ij}
\ee
with obvious meaning of the symbols. Since the Cooper pairs undergo Bose-Einstein
condensation their total momentum is zero. Assuming that also the spin is zero, we get {\em simple pairing}\/: the fermions of the pair have opposite momenta and opposite spins, so that the Fourier transform $\tilde{\varphi}$ of the pair wave function can be written as a function of a unique variable
\be
\varphi_k := \tilde{\varphi}(k,+;-k,-) \,.
\ee
Here, and in the following, according to the standard notation, the sum over $k$ implies also the sum over spins according to the
previous identification. So that, if the probability of finding a fermion with momentum $k$ is
\be
\rho_k = \frac{ \varphi_k^2}{ 1+ \varphi_k^2}\,,
\ee
the condition of having $N$ fermions is
\be
\sum_k \rho_k =N\,. \label{fernum}
\ee

To minimize the expectation value of the Hamiltonian under this condition,  a chemical potential ${\mu}$ is introduced, so that one has to minimize the quantity 
\be
{\mathcal E} = \langle H_N - \mu N_{op} \rangle
\ee
where $N_{op}$ is the fermion number operator. 
The above expectation value, whose evaluation~\cite{Blat} is far from trivial is
\begin{eqnarray}
{\mathcal E} &=&  \sum_k (e_k - \mu) \, \rho_k  + \frac{1 }{2} \sum_{k, l} ( \langle k,l|V|k,l \rangle -  \langle k,l|V|l, k \rangle )\, \rho_k \rho_l
\nonumber\\
&  & + \frac{1 }{2} \sum_{k, l} \langle  k, -k |V|l ,- l\rangle\, \psi_k \psi_l  + O\left( \frac{1}{N}\right)\label{vac energy}
\end{eqnarray}
where
\be
\epsilon_k := \frac{k^2}{2M}\,, \qquad \psi_k := \frac{ \varphi_k}{ 1+ \varphi_k^2} \,.
\ee
Terms of order $O\left( \frac{1}{N}\right) $ have been neglected. They all respect fermion number conservation, and can be disregarded 
below the critical temperature, but are essential in the determination of the critical properties of the transition to normal state.

The first term is the contribution of the kinetic energy, the second of the density-density interaction, and the third of the interaction of fermions in one and the same pair. Because the density-density interaction is almost the same in the normal and in the superconducting state, to simplify the calculation it is accounted  for by a renormalization of the chemical potential
\be
\mu_{\rm eff} = \mu - \frac{1 }{2} \sum_{l} \left( \langle k,l|V|k,l \rangle -  \langle k,l|V|l, k \rangle \right)\, \rho_l \,.
\ee
Variation with respect to $\varphi$ at constant $\mu_{eff}$ gives
\be
(\epsilon_k- \mu_{\rm eff})\, \varphi_k + \frac{1}{2} (\varphi_k^2 - 1)\, \Delta_k =0 
\ee
where 
\be
\Delta_k = -  \frac{1 }{2} \sum_{l}  \left( \langle  k, -k |V|l ,- l\rangle +  \langle  l, -l |V|k, -k \rangle \right)\, \psi_l
\ee
is the gap function. The solutions 
\be
\varphi_k= - \frac{\epsilon_k - \mu_{\rm eff} }{\Delta_k} \pm \sqrt{ 1 + \left(\frac{\epsilon_k - \mu_{\rm eff}}{\Delta_k}\right)^2 }
\ee
 inserted in the definition of the gap function give the gap equation
\be
\Delta_k = \mp \sum_l  \frac{ \langle  k, -k |V| l ,- l\rangle }{2 \left[ 1+ \left(\frac{\epsilon_{l} - \mu_{\rm eff}}{\Delta_{l}}\right)^2  \right]}\,\Delta_l \, .
\ee
In order to get a close solution some approximations are needed. 
First, a separable form is assumed for the matrix elements of the potential, that is
\be
 \langle  k, -k |V|l ,- l\rangle \approx 
 \begin{cases} - \frac{\gamma^2}{L^3} & \mbox{for\, } \, |\epsilon_k -\mu_{\rm eff}| < \omega \,; \\
 0 & \mbox{otherwise.}
 \end{cases}
\ee
 $L^3$ is the volume of the system. This is a crude but reasonable approximation of the electron-electron 
 interaction due to phonon exchange. Second, since the contribution to the integral comes essentially from $e_k \approx \mu_{\rm eff}
 \approx \frac{k_F^2}{2M}$, where $k_F$ is the Fermi momentum, one can set $\Delta_k \approx \Delta_{k_F}
 = \Delta$. Then the solution of the gap equation is
\be
\Delta \approx 2 \, \omega \exp \left( -  \frac{Mk_F}{2 \pi^2\gamma^2}\right) \,.
\ee
It is important to observe, in connection with our problem of including Cooper pairs and unpaired fermions at the same time, that, for 
$|\epsilon_k- \mu_{\rm eff} | > \omega$, we assume $\Delta_k =0$, but $\varphi_k$ need not to vanish. Actually in general $\varphi_k$ must not vanish for 
$ \frac{p_k^2}{2m}< \mu_{\rm eff}- \omega$ in order to fulfill the condition~(\ref{fernum}) on fermion number.

We conclude  this subsection by discussing the effect of the coupling to a magnetic field. The Cooper pair structure function depends on the applied gauge field, and the interaction can be computed in a perturbation series, that is
\be
\varphi = \varphi_0 + e\, \varphi_1({\bf A}) +\,O(e^2) \,.  \label{mag suscett}
\ee
The first correction $\varphi_1({\bf A})$ ($e$ being the electric charge and $\bf A$ the vector potential) describes the Cooper pair magnetic susceptibility, and is essential in the explanation of the Meissner effect. 
Now the electromagnetic coupling  is small with respect to the phonon coupling, which binds the electrons in a Cooper pair~\cite{Blat}, and therefore the perturbation expansion is justified.  In the case of QCD  it is, instead, the gauge interaction which binds quarks into diquarks, and therefore, in general, the dependence of the diquark structure function on the gauge fields cannot be neglected.

\subsection{Bogoliubov transformations}

The Bogoliubov transformations corresponding to simple pairing in standard notations are
\be
\alpha_k^{\dagger}= u_k  \, c_k^{\dagger} -  v_k \, c_{-k} \,, \qquad \label{Bogo-manybody}
\alpha_k= u_k  c_k -  v_k c_{-k}^{\dagger} 
\ee
where $c_k^{\dagger}, c_k$ are creation-annihilation operators of the fermions in the system and the parameters $u,v$, not to be confused with the upper and lower spinor components,  must satisfy the normalization conditions
\be
u_k^2 +v_k^2 =1 \,. 
\ee
The transformed Hamiltonian is
\be
H' - \mu N_{op}= {\mathcal E} + H_{11} + H_{20} +H_{int}
\ee
where $N_{op}$ is the fermion number operator
\begin{align}
{\mathcal E} = & \sum_k ( e_k -\mu)  v_k^2 + \frac{1}{2}\sum_{kl} u_k u_lv_k v_l \, \langle k,-k|V| l,-l \rangle \\
H_{11} = & \sum_k \left[ (e_k -\mu)  (u_k^2  -  v_k^2)- 2 \sum_{l} u_k u_l v_k v_l 
 \, \langle k,-k|V| l,-l \rangle \right]  \alpha_k^{\dagger} \alpha_k \\
H_{20} = & \sum_k \left[ (e_k - \mu ) u_k\, v_k + \frac{1}{2} (u_k^2 - v_k^2 )
\sum_l u_lv_l \, \langle k,-k|V| l,-l \rangle\right]\\
 &  \times  ( \alpha_k^{\dagger}\alpha_{-k}^{\dagger} + \alpha_{-k} \alpha_{k} ) \nonumber
\end{align}
and the density-density interaction has been neglected. These terms have a close correspondence with those of the transformed action~(\ref{SF}).
We have not written  $H_{int}$. We only mention that it contains monomials of operators of power higher than 2 which do not conserve
fermion number  but are    of order $\frac{1}{N}$ ( essential in the study of the phase transition to normal state).
 If we set
\be
v_k =  \frac{f_k }{\sqrt{1+f_k^2}}       \,, \qquad u_k=\frac{1}{\sqrt{1+f_k^2}}    \label{Bogoparameters}
\ee
we see that the vacuum energy ${\mathcal E}$ is identical to that found in the quasichemical equilibrium theory. At its minimum, $H_{20}=0$,
in perfect analogy with the results we got in the relativistic case, and 
\be
H_{11}= \sum_k \sqrt{ (e_k - \mu)^2 + \Delta^2} \, \alpha_k^{\dagger} \alpha_k 
\ee
so that the Bogoliubov-Valatin method gives directly also the spectrum of quasiparticles which in the quasichemical equilibrium theory has to found separately. Introducing the parametrization~(\ref{Bogoparameters}) into the definitions~(\ref{Bogo-manybody}) we can recognize 
the form of our relativistic transformations. If we make the Bogoliubov transformation time-dependent, we can conserve fermion number by the help of compensating fields. The development of this approach for many-body systems can be found in
\cite{Palu2}.

\section{Derivation of the diquark effective action}

\subsection{Pfaffians}

We first need to recall some basic facts about pfaffians. The interested reader can find a discussion about the properties of pfaffians and their relation to the Gaussian Berezin integrals in the detailed appendices of~\cite{CSS}, together with similar properties of determinants, permanents and hafnians.

Let $A = (A_{ij})_{i,j=1}^{2m}$ be a be a $2m \times 2m$ {\em antisymmetric}\/ matrix.
We define the {\em pfaffian} of $A$ by
\be
   \pf A  \;=\;  \frac{1}{2^m\, m!} \,
                \sum_{\sigma \in {\mathcal{S}}_{2m}}  \sgn(\sigma) \,
                  A_{\sigma(1) \sigma(2)} \cdots A_{\sigma(2m-1) \sigma(2m)}
 \label{def.pfA}
\ee
where ${\mathcal{S}}_{2m}$ is the {\em symmetric group} of $2m$ elements, and $\sgn(\sigma)$ is the sign of the permutations $\sigma$.
Then
\be
(\pf A)^2 = \det A \label{pf2}
\ee
and for any $2m \times 2m$ matrix $X$
\be
\pf \left(X A X^T \right) = (\det X) (\pf A) \, . \label{pfdet}
\ee
If $A$ is invertible
\be
\pf \left( A^{-T} \right) = (\pf A)^{-1}\, .
\ee
Consider a partitioned matrix of the form
\be
   M \;=\; \left( \begin{array}{c|c}
                    A & B \\
                    \hline
                    -B^{\rm T} & D
               \end{array} \right)
\ee
where $A,B,D$ are matrices of sizes
$2m \times 2m$, $2m \times 2n$ and $2n \times 2n$, respectively,
with elements in a commutative ring with identity,
and $A$ and $D$ are antisymmetric.
If $A$ is invertible, then
\be
\pf M = (\pf A) \pf(D + B^{\rm T} A^{-1} B) \, . \label{pfpm1}
\ee
If $D$ is invertible, then
\be
\pf M = (\pf D) \pf(A + B D^{-1} B^{\rm T}) \, . \label{pfpm2}
\ee

Now let $\chi_1, \dots, \chi_n$ be the generators of a Grassmann algebra and $A$ be an antisymmetric $n \times n$ matrix. Then the Gaussian Berezin integral provides a representation for the pfaffian:
\be
\int \! d\chi_1 \cdots d\chi_n \exp \left( \frac{1}{2} \sum_{i, j=1}^n \chi_i A_{ij} \chi_j \right) = 
\begin{cases}
\pf A  & \hbox{if $n$ is even}\\
0        & \hbox{if $n$ id odd}\, .
\end{cases}
\ee

\subsection{Evaluation of the transfer matrix}

In this section we refer to the fermion transfer matrix at non-zero chemical potential $\mu$, in an arbitrary gauge, as defined in~\reff{calt}, with the notations given in~\reff{hatn} and~\reff{transfer}.

Let $| u v \rangle$ be the coherent state associated to the fermion operators $\hat{u}$ and $\hat{v}$, that is
\be
| u v \rangle = \exp(-u\hat{u}^\dagger - v\hat{v}^\dagger) |0\rangle 
\ee
where $u$ and $v$ are Grassmann variables.
We shall make use of the completeness relation
\be
1 = \int \! du du^* dv dv^* \frac{| u v \rangle \langle u v | }{\langle u v | u v \rangle}
\ee
by the help of the Berezin integration on Grassmann variables.

The scalar product of two states is
\be
\langle u_1 v_1| u_2 v_2 \rangle = \exp \left(u_1^* u_2 + v_1^* v_2  \right) \, .
\ee

The evaluation of the matrix element of the transfer matrix between coherent states was already performed in~\cite{CPV} with the result
\begin{multline}
\langle u_t v_t|{\mathcal T}_{t,t+1} |u_{t+1}  v_{t+1} \rangle =  \\
\exp \left( u_t^* N_t^\dagger v_t^* + v_{t+1} N_{t+1} u_{t+1} + u_t^* U_{0,t} e^{2 \mu} u_{t+1} + v_t^* U_{0,t}^* e^{-2 \mu} v_{t+1}\right) \, .
\end{multline}

Here we are interested in the evaluation of the matrix element
\be
{\cal I} :=  \langle {\mathcal D}_t, {\mathcal F}_t | {\mathcal T}_{t,t+1} | {\mathcal D}_{t+1}, {\mathcal F}_{t+1} \rangle \, .
\ee
Our procedure goes through the introduction of two complete sets of coherent states
\begin{multline}
{\cal I} = \int \! du_t du_t^* dv_t dv_t^* du_{t+1} du_{t+1}^* dv_{t+1} dv_{t+1}^* \\
\frac{ \langle {\mathcal D}_t, {\mathcal F}_t | u_t v_t \rangle}{\langle u_t v_t | u_t v_t \rangle} \langle u_t v_t|{\mathcal T}_{t,t+1} |u_{t+1}  v_{t+1} \rangle \frac{ \langle u_{t+1} v_{t+1}  | {\mathcal D}_{t+1}, {\mathcal F}_{t+1} \rangle}{\langle u_{t+1} v_{t+1} | u_{t+1} v_{t+1} \rangle}\, . 
\end{multline}
Also to compute the matrix element $\langle u_1 v_1| {\mathcal D}, {\mathcal F} \rangle$ we insert a complete set of coherent states as follows
\be
\langle u_1 v_1| {\mathcal D}, {\mathcal F} \rangle = \int \! du_2 du_2^* dv_2 dv_2^* \, \frac{ \langle u_1 v_1| \exp (  \hat{ {\mathcal D}}^{\dagger}/2 ) | u_2 v_2 \rangle \langle u_2 v_2| \exp ( \hat{ {\mathcal F}}^\dagger ) | 0  \rangle}{\langle u_2 v_2 | u_2 v_2 \rangle}\, .
\ee

According to our definitions~\reff{hatD} and~\reff{Bogoliubov}
\be
\hat{ {\mathcal D}}^{\dagger}  = 
\hat{u}^\dagger  R^\frac{1}{2} {\mathcal D}^{\dagger} R^{* \frac{1}{2}} \hat{u}^\dagger + \hat{v} {\mathcal F} R^\frac{1}{2} {\mathcal D}^{\dagger} R^{* \frac{1}{2}}  {\mathcal F}^T \hat{v} - 2  \hat{u}^\dagger R^\frac{1}{2} {\mathcal D}^{\dagger} R^{* \frac{1}{2}} {\mathcal F}^T \hat{v}
\ee
This means that
\begin{multline}
\langle u_1 v_1| \exp (  \hat{ {\mathcal D}}^{\dagger}/2 ) | u_2 v_2 \rangle =  \langle u_1 v_1| u_2 v_2 \rangle \\ 
\times \exp \left( \frac{1}{2} u_1^*  R^\frac{1}{2} {\mathcal D}^{\dagger} R^{* \frac{1}{2}} u_1^* + \frac{1}{2} v_2 {\mathcal F} R^\frac{1}{2} {\mathcal D}^{\dagger} R^{* \frac{1}{2}}  {\mathcal F}^T v_2 - u_1^*  R^\frac{1}{2} {\mathcal D}^{\dagger} R^{* \frac{1}{2}} {\mathcal F}^T v_2 \right) \, .
\end{multline}
Similarly
\be
\langle u_2 v_2| \exp ( \hat{ {\mathcal F}}^\dagger ) | 0  \rangle = \exp \left( u_2^* {\mathcal F}^\dagger v_2^* \right) \, .
\ee
Therefore
\begin{multline}
\langle u_1 v_1| {\mathcal D}, {\mathcal F} \rangle = \exp \left( \frac{1}{2} u_1^*  R^\frac{1}{2} {\mathcal D}^{\dagger} R^{* \frac{1}{2}} u_1^* \right) \, \int \! du_2 du_2^* dv_2 dv_2^* \, \exp \left( u_2^* {\mathcal F}^\dagger v_2^* \right) \\
\times \exp\left( \frac{1}{2} v_2 {\mathcal F} R^\frac{1}{2} {\mathcal D}^{\dagger} R^{* \frac{1}{2}}  {\mathcal F}^T v_2 - u_1^*  R^\frac{1}{2} {\mathcal D}^{\dagger} R^{* \frac{1}{2}} {\mathcal F}^T v_2 + u_1^* u_2 + v_1^* v_2 - u_2^* u_2 - v_2^* v_2\right) \, .
\end{multline}
The integrations on the variables $u_2$ and $v_2^*$ produce, respectively, the constraints $u_2^* = u_1^*$ and $v_2 = u_1^* {\mathcal F}^\dagger$, and we arrive at the result
\be
\langle u_1 v_1| {\mathcal D}, {\mathcal F} \rangle =  \exp \left( \frac{1}{2} u_1^*  R^{-\frac{1}{2}} {\mathcal D}^{\dagger} (R^{-T})^\frac{1}{2} u_1^* + u_1^* {\mathcal F}^\dagger v_1^* \right)\, .
\ee
By using all these intermediate steps we get
\begin{multline}
{\cal I} = \int \! du_t du_t^* dv_t dv_t^* du_{t+1} du_{t+1}^* dv_{t+1} dv_{t+1}^* \\
\times \exp \left( \frac{1}{2} u_{t}  (R_{t}^{-T})^\frac{1}{2} {\mathcal D}_{t} R_{t}^{-\frac{1}{2}} u_{t} + v_{t} {\mathcal F}_{t} u_{t} - u_t^* u_t  - v_t^* v_t \right)\\
\times \exp \left( u_t^* N_t^\dagger v_t^* + v_{t+1} N_{t+1} u_{t+1} + u_t^* U_{0,t} e^{2 \mu} u_{t+1} + v_t^* U_{0,t}^* e^{-2 \mu} v_{t+1}\right)  \\
\times \exp \left( \frac{1}{2} u_{t+1}^*  R_{t+1}^{-\frac{1}{2}} {\mathcal D}_{t+1}^{\dagger} (R_{t+1}^{-T})^\frac{1}{2} u_{t+1}^* + u_{t+1}^* {\mathcal F}_{t+1}^\dagger v_{t+1}^*  - u_{t+1}^* u_{t+1}   - v_{t+1}^* v_{t+1} \right)\, .
\end{multline}
The integrations on the variables $v_{t+1}^*$ and $v_t$ produce, respectively, the constraints $v_{t+1} = - u_{t+1}^* {\mathcal F}_{t+1}^\dagger$ and $v_t^*  = - {\mathcal F}_t u_t$, and, using the definition of ${\mathcal F}_{N,t}$ given in~\reff{FN}, we arrive at the expression
\begin{multline}
{\cal I} = \\
\int \! du_t du_t^*  du_{t+1} du_{t+1}^*  \exp \left( \frac{1}{2} u_{t}  (R_{t}^{-T})^\frac{1}{2} {\mathcal D}_{t} R_{t}^{-\frac{1}{2}} u_{t}+ \frac{1}{2} u_{t+1}^*  R_{t+1}^{-\frac{1}{2}} {\mathcal D}_{t+1}^{\dagger} (R_{t+1}^{-T})^\frac{1}{2} u_{t+1}^*   \right) \\
\times \exp \left(  - u_t^* {\mathcal F}_{N,t} u_t - u_{t+1}^* {\mathcal F}_{N,t+1}^\dagger u_{t+1}  + u_t^* U_{0,t} e^{2 \mu} u_{t+1} +  u_t {\mathcal F}_t^T U_{0,t}^* e^{-2 \mu} {\mathcal F}_{t+1}^* u_{t+1}^*  \right)\, .
\end{multline}
As the next step we perform the Berezin integrations on $u_{t+1}$ and $u_{t+1}^*$ to get
\begin{multline}
{\cal I} = 
\det({\mathcal F}_{N,t+1}^\dagger) \, \int \! du_t du_t^* \exp \left(   \frac{1}{2} u_{t}  (R_{t}^{-T})^\frac{1}{2} {\mathcal D}_{t} R_{t}^{-\frac{1}{2}} u_{t}   \right)  \\
\times \exp \left[ - u_t^* \left( {\mathcal F}_{N,t}  +U_{0,t} ({\mathcal F}_{N,t+1}^\dagger)^{-1} {\mathcal F}_{t+1}^\dagger U_{0,t}^\dagger  {\mathcal F}_t \right) u_t  \right]\\
\times \exp \left(  \frac{1}{2} u_{t}^* U_{0,t} e^{2\mu} ({\mathcal F}_{N,t+1}^\dagger)^{-1} R_{t+1}^{-\frac{1}{2}} {\mathcal D}_{t+1}^{\dagger} (R_{t+1}^{-T})^\frac{1}{2} ({\mathcal F}_{N,t+1}^\dagger)^{-T} e^{2\mu} U_{0,t}^T u_{t}^* \right)\, .
\end{multline}
These final integrals produce the pfaffian of a partitioned matrix and can be computed by using the expressions~\reff{pfpm1} or~\reff{pfpm2} to get
\begin{multline}
{\cal I} =  \det({\mathcal F}_{N,t+1}^\dagger) \, \pf \left[(R_{t}^{-T})^\frac{1}{2} {\mathcal D}_{t} R_{t}^{-\frac{1}{2}} \right] \, \\
\times \pf\left\{ U_{0,t} e^{2\mu} ({\mathcal F}_{N,t+1}^\dagger)^{-1} R_{t+1}^{-\frac{1}{2}} {\mathcal D}_{t+1}^{\dagger} (R_{t+1}^{-T})^\frac{1}{2} ({\mathcal F}_{N,t+1}^\dagger)^{-T} e^{2\mu} U_{0,t}^T +\right. \\ 
 [{\mathcal F}_{N,t}  +U_{0,t} ({\mathcal F}_{N,t+1}^\dagger)^{-1} {\mathcal F}_{t+1}^\dagger U_{0,t}^\dagger  {\mathcal F}_t][(R_{t}^{-T})^\frac{1}{2} {\mathcal D}_{t} R_{t}^{-\frac{1}{2}} ]^{-1} \\
\left. \times [{\mathcal F}_{N,t}  +U_{0,t} ({\mathcal F}_{N,t+1}^\dagger)^{-1} {\mathcal F}_{t+1}^\dagger U_{0,t}^\dagger  {\mathcal F}_t]^T \right\}
\end{multline}
where the last pfaffian can be re-written, by using formula~\reff{pfdet}, as
\begin{multline}
{\cal I} =  \det({\mathcal F}_{N,t+1}^\dagger) \, \pf \left[(R_{t}^{-T})^\frac{1}{2} {\mathcal D}_{t} R_{t}^{-\frac{1}{2}} \right] \, 
\det[{\mathcal F}_{N,t}  +U_{0,t} ({\mathcal F}_{N,t+1}^\dagger)^{-1} {\mathcal F}_{t+1}^\dagger U_{0,t}^\dagger  {\mathcal F}_t] \, \\
\times \pf \left\{ [ (R_{t}^{-T})^\frac{1}{2} {\mathcal D}_{t} R_{t}^{-\frac{1}{2}} ]^{-1} + 
[{\mathcal F}_{N,t}  +U_{0,t} ({\mathcal F}_{N,t+1}^\dagger)^{-1} {\mathcal F}_{t+1}^\dagger U_{0,t}^\dagger  {\mathcal F}_t]^{-1} \right.\\
U_{0,t} e^{4\mu} ({\mathcal F}_{N,t+1}^\dagger)^{-1} R_{t+1}^{-\frac{1}{2}} {\mathcal D}_{t+1}^{\dagger} (R_{t+1}^{-T})^\frac{1}{2} ({\mathcal F}_{N,t+1}^\dagger)^{-T}  U_{0,t}^T \\
\left. \times [{\mathcal F}_{N,t}  +U_{0,t} ({\mathcal F}_{N,t+1}^\dagger)^{-1} {\mathcal F}_{t+1}^\dagger U_{0,t}^\dagger  {\mathcal F}_t]^{-T} \right\}\, . \label{inmezzo}
\end{multline}
Remark that $\det({\mathcal F}_{N,t+1}^\dagger) =  \det({\mathcal F}_{N,t+1}^\dagger U_{0,t}^\dagger)$ so that the product of  this determinant with the other one appearing  in~\reff{inmezzo} can be written as the determinant of the product, which is exactly $E_{t+1,t}$ according to~\reff{E}, so that
\begin{multline}
{\cal I} =  \det( E_{t+1,t} ) \, 
\pf \left[(R_{t}^{-T})^\frac{1}{2} {\mathcal D}_{t} R_{t}^{-\frac{1}{2}} \right] \, \\
\times  \pf \left\{ \left[ (R_{t}^{-T})^\frac{1}{2} {\mathcal D}_{t} R_{t}^{-\frac{1}{2}} \right]^{-1} + 
 e^{4 \mu} E_{t+1,t}^{-1} R_{t+1}^{-\frac{1}{2}} {\mathcal D}_{t+1}^{\dagger} (R_{t+1}^{-T})^\frac{1}{2}
E_{t+1,t}^{-T}\right\}\, .
\end{multline}
By  using the relation~\reff{newdefQ} and the formula~\reff{pfdet}, we obtain, at the end, the expression
\be
{\cal I} =  \det( E_{t+1,t} ) \pf ( {\mathcal D}_{t} ) \pf ( {\mathcal D}_{t}^{-1} + e^{4 \mu} Q_{t,t+1}^{-1} {\mathcal D}_{t+1}^{\dagger} Q_{t+1,t}^{-T} )\, .
\ee

We shall also need the normalization factor
\begin{multline}
\langle {\mathcal D}, {\mathcal F} | {\mathcal D}, {\mathcal F} \rangle = \int \! du du^* dv dv^* \exp \left( - u^* u - v^* v 
+ u^* {\mathcal F}^\dagger v^*  + v {\mathcal F} u\right) \\
\times \exp\left( \frac{1}{2} u^* R^{-\frac{1}{2}} {\mathcal D}^\dagger (R^{-T})^\frac{1}{2} u^* + \frac{1}{2} u (R^{-T})^{\frac{1}{2}} {\mathcal D}^\dagger R^{-\frac{1}{2}} u \right)\, ,
\end{multline}
which after the integration on $v^*$ and $v$ and a rescaling of the variables becomes
\begin{align}
\langle {\mathcal D}, {\mathcal F} | {\mathcal D}, {\mathcal F} \rangle =  & \,
(\det R)^{-1}
\int \! du du^*  \exp \left( - u^*  u + \frac{1}{2}  u^* {\mathcal D}^\dagger  u^* + \frac{1}{2} u  {\mathcal D}^\dagger  u \right) \\
= & \, (\det R)^{-1} \pf ({\mathcal D}) \pf( {\mathcal D}^{-1} + {\mathcal D}^\dagger )\, .
\end{align}
In conclusion
\begin{align}
\exp  (- S_{bo} ) = & \, \prod_{t=0}^{L_0/2-1} \frac{\langle {\mathcal D}_t, {\mathcal F}_t | {\mathcal T}_{t,t+1} | {\mathcal D}_{t+1}, {\mathcal F}_{t+1} \rangle }{\langle {\mathcal D}_t, {\mathcal F}_t | {\mathcal D}_t, {\mathcal F}_t \rangle } \\
= & \, \prod_{t=0}^{L_0/2-1} \frac{\det( E_{t+1,t} ) \pf ( {\mathcal D}_{t} ) \pf ( {\mathcal D}_{t}^{-1} + e^{4 \mu} Q_{t,t+1}^{-1} {\mathcal D}_{t+1}^{\dagger} Q_{t+1,t}^{-T} )}{(\det R)_t^{-1} \pf ({\mathcal D}_t) \pf( {\mathcal D}_t^{-1} + {\mathcal D}_t^\dagger )} \\
= & \, \prod_{t=0}^{L_0/2-1}  \det( Q_{t+1,t} ) 
\left[\frac{\det ( 1 + e^{4 \mu} {\mathcal D}_{t} Q_{t,t+1}^{-1} {\mathcal D}_{t+1}^{\dagger} Q_{t+1,t}^{-T} )}{\det (1 + {\mathcal D}_{t} {\mathcal D}_{t}^{\dagger} )}\right]^\frac{1}{2} \, .
\end{align}
From this expression, as $S_{me}$ is the part of $S_{bo}$ at ${\mathcal D} = 0$ and $S_{dq}$ the rest, we easily derive that
\begin{align}
S_{me} = & \, - \sum_{t=0}^{L_0/2-1} \tr_+ \ln Q_{t+1,t} \\
S_{dq} = & \, \frac{1}{2} \sum_{t=0}^{L_0/2-1} \tr_+ \left[ \ln ( 1 +  {\mathcal D}_{t} {\mathcal D}_{t}^{\dagger} ) - \ln ( 1 + e^{4 \mu} {\mathcal D}_{t} Q_{t,t+1}^{-1} {\mathcal D}_{t+1}^{\dagger} Q_{t+1,t}^{-T} )\right]\, . 
\end{align}

\end{document}